\documentclass[ras]{agutex}

\usepackage[dvipdfm]{graphicx}

\authorrunninghead{ROGERS and BOWMAN}

\titlerunninghead{Absolute calibration ver 4 Oct 11}

\begin{document}

\title{Absolute calibration of a wideband antenna and spectrometer for sky noise spectral index measurements}

\authors{Alan E. E. Rogers,\altaffilmark{1} and Judd D. Bowman,\altaffilmark{2}}

\altaffiltext{1}{MIT Haystack Observatory, Westford, Massachusetts, USA.}

\altaffiltext{2}{Arizona State University, Tempe, Arizona, USA.}

\begin{abstract}
 A new method of absolute calibration of sky noise temperature using a 3-position switched spectrometer,
measurements of antenna and low noise amplifier impedance with a vector network analyzer, and ancillary measurements
of the amplifier noise waves is described. The details of the method and its application to accurate wideband
measurements of the spectral index of the sky noise are described and compared with other methods.
\end{abstract}

\begin{article}

\section{Introduction}
 Accurately calibrated sky noise observations are needed to measure the spectral index of the diffuse
radio background which is important in determining the emission mechanisms in the Galactic and extragalactic background.
Theoretical models of the early universe predict small deviations in the spectrum of the radio sky due to emission and absorption
of neutral hydrogen at redshifts which place the 21-cm line in the 50 - 200 MHz range. In addition to the applications in radio astronomy
accurately calibrated sky noise measurements are used to measure the absorption of radio signals in the ionosphere using riometers
[{\textit  {Hargreaves and Detrick}}, 2002].
Also accurate calibration has applications in antenna pattern measurements and spectrum monitoring.
  The measurement of the antenna temperature is affected by the antenna impedance as any mismatch to the receiver results
in a fraction of the sky noise power, $T_{sky}$, being reflected back into the sky so that the measured noise temperature from the
antenna, $T_{ant}$, is

\begin{equation}
 T_{ant} = T_{sky}(1 -  |\Gamma|^2)
\end{equation}  
where $\Gamma$ is the reflection coefficient. In addition outgoing noise from the receiver is reflected back to the receiver
by the mismatch. The "mismatch" can be corrected by using a mechanical tuner to adjust the antenna impedance to obtain the same impedance
as the calibrated noise source used to calibrate the receiver. Under this condition any power lost due to the Low Noise Amplifier
(LNA) having a different impedance
is the same for the antenna as it is for the noise source.
Several accurately calibrated measurements of the sky noise spectral index using this method were published in the 1960s and 1970s
[{\textit  {Pauliny-Toth and Shakeshaft}}, 1962; {\textit  {Turtle et al.}}, 1962]. However this method is limited to a narrow frequency range
and requires constructing scaled antennas to cover a wide frequency range. 
With the development of broadband antennas, amplifiers and spectrometers new methods
are needed for automated measurements over wide frequency range.

\section{50 ohm measurements at a fixed reference plane}

  A vector network analyzer (VNA) provides an accurate measurement of the reflection coefficient referenced to the standard 50 ohm
impedance and consequently it is convenient to use a VNA to measure both the antenna and receiver input reflection coefficient
at the fixed reference plane defined by the 50 ohm connection between the receiver and the antenna. 
In this case, with some algebraic manipulation equation (1) becomes

\begin{equation}
 T_{sky}(1 -  |\Gamma|^2) = T_{sky}(1 -  |\Gamma_a|^2)|F|^2
\end{equation}
where

\begin{equation}
 \Gamma = \frac{(Z_a - Z_l^*)}{(Z_a + Z_l)}
\end{equation}

\begin{equation}
 F = \frac{(1 -  |\Gamma_l|^2)^{1/2}}{(1 - \Gamma_a\Gamma_l)}
\end{equation}

\begin{equation}
 \Gamma_a = \frac{(Z_a - 50)}{(Z_a + 50)}
\end{equation}

\begin{equation}
 \Gamma_l = \frac{(Z_l - 50)}{(Z_l + 50)}
\end{equation}
$Z_a,Z_l$ and $\Gamma_a,\Gamma_l$ are the impedance and reflection coefficients of the antenna and receiver respectively.
Part of the complex factor F can also be viewed as the sum of the noise waves back and forth from the antenna to the receiver
since the sum is a polylogarithmic series which converges when $|\Gamma_a\Gamma_l| < 1$ and has an exact solution

\begin{equation}
 \sum\limits_{i=0}^\infty (\Gamma_a\Gamma_l)^i = \frac{1}{(1-\Gamma_a\Gamma_l)}.
\end{equation}

We can add the receiver LNA noise waves to equation (2) to obtain the noise
temperature, $T_{rec}$, from the LNA

\begin{eqnarray}
 T_{rec} = T_{sky}(1 -  |\Gamma_a|^2)|F|^2 + T_u|\Gamma_a|^2|F|^2 \nonumber \\
  + (T_ccos(\phi)+T_ssin(\phi))|\Gamma_a||F| + T_0
\end{eqnarray}
where $T_u$ is the uncorrelated portion of the LNA noise reflected from the antenna, $T_c$ and $T_s$ are the cosine and sine components of the correlated noise
from the LNA reflected by the antenna and $T_0$ is the portion of the noise which is independent of the LNA input. $\phi$ is the phase of $\Gamma_aF$. Equation (8)
is equivalent to the noise wave formulation of {\textit  {Meys} [1978]} if the "reference" impedance is that of the LNA in which case $F=1$.

\section{Receiver calibration}
 
  A three position switch is used to provide the receiver calibration. This switch sequentially connects the LNA
to the antenna, a 50 ohm ambient load and a 50 ohm calibrated noise source. The power in each position, $P_{ant},P_{load},P_{cal}$ is

\begin{equation} 
 P_{ant} = gT_{rec}
\end{equation}

\begin{equation} 
 P_{load} = g(GT_{amb} + T_0)
\end{equation}

\begin{equation} 
 P_{cal} = g(G(T_{amb}+T_{cal}) + T_0)
\end{equation}
where

\begin{equation} 
 G = 1 - |\Gamma_l|^2
\end{equation}
and $g$ is the receiver gain, $Tamb$ is the ambient temperature, $T_{cal}$ is the excess noise temperature of the calibration
noise. The calibrated receiver output, $T_{3p}$, is

\begin{eqnarray} 
 T_{3p} & = & \frac{T_{cal}(P_{ant}-P_{load})}{(P_{cal}-P_{load})} + T_{amb} \nonumber \\
        & = & T_{sky}(1-|\Gamma_a|^2)|F|^2G^{-1} \nonumber \\
        & + & T_u|\Gamma_a|^2|F|^2 G^{-1} \nonumber \\
        & + & (T_ccos(\phi)+T_ssin(\phi))|\Gamma_a||F|G^{-1} 
\end{eqnarray}

\section{Measurement of LNA noise wave parameters}

  The LNA noise wave parameters, $T_u$,$T_c$ and $T_s$ can be measured by connecting an open (or shorted)
low loss cable to the calibrated 3-position switched spectrometer in place of the antenna. The cable acts just like
an antenna looking at an isotropic sky with temperature equal to the cable's physical temperature. 
The cable reflection coefficient, $\Gamma_c$, should be 
approximately

\begin{equation} 
 \Gamma_c = L_c exp(-i\omega\tau)
\end{equation}
where $\tau$ is the 2-way delay, $\omega$ is the frequency in radians/s and $L_c$ is the cable loss factor 

\begin{equation} 
 L_c = 10^{-lc/10}
\end{equation}
where $lc$ is the cable loss in dB. For highest accuracy
the open cable's reflection coefficient should be measured with the VNA since most cables are not exactly 50 ohms and the
reflection coefficient is more complex than the expression above. With the open cable connected the calibrated receiver output is given by equation
(13) by substitution of $T_{sky}$ and $\Gamma_a$ with $T_{amb}$ and $\Gamma_c$. Since the LNA noise parameters change gradually with
frequency the rapid change of phase, $\phi$, afforded by a cable of sufficient length allows the noise wave parameters to be separately estimated 
via a least squares fit to the functions $|\Gamma_a|^2|F|^2G^{-1}$, $cos(\phi)|\Gamma_a||F|G^{-1}$ and $sin(\phi)|\Gamma_a||F|G^{-1}$.

\section{Corrections for antenna losses}

  If the antenna is lossless the calibrated sky noise temperature can be obtained from equation (13)

\begin{eqnarray} 
 T_{sky} & = & [T_{3p} - T_u|\Gamma_a|^2|F|^2G^{-1} \nonumber \\
   & - & (T_ccos(\phi)+T_ssin(\phi))|\Gamma_a||F|G^{-1}] \nonumber \\ 
  & \times & [(1-|\Gamma_a|^2)|F|^2G^{-1}]^{-1}
\end{eqnarray}
noting that when $\Gamma_a=0$, $T_{sky}=T_{3p}$.

If the antenna loss factor is $L$ the corrected sky temperature, $T_{csky}$, is

\begin{equation} 
 T_{csky} = (T_{usky} - T_{amb}(L-1))/L
\end{equation}
where $T_{usky}$ is the uncorrected sky temperature and

\begin{equation} 
 L = 10^{-l/10}
\end{equation}
where $l$ is the loss in dB. For a typical antenna the loss is made of several components which include ground loss, 
resistive loss, transmission line loss and balun loss. The ground loss factor is one minus the fraction of the antenna
pattern which receives thermal radiation from the ground. The resistive loss is one minus the fraction of the real part
of the antenna impedance that comes from the conductor resistance including the skin-effect. If a transmission line 
and balun are part of the antenna the loss can be incorporated into a 50 ohm attenuation in equation (8) which becomes

\begin{eqnarray}
 T_{rec} & = & T_{sky}(1 -  |\Gamma_a|^2)|F|^2L \nonumber \\ 
  & + & T_{amb}((1-L)L|\Gamma_a|^2+1-L) \nonumber \\
  & + & T_u|\Gamma_a|^2|F|^2L^2 \nonumber \\
  & + & (T_ccos(\phi)+T_ssin(\phi)|\Gamma_a||F|L + T_0
\end{eqnarray}
where 

\begin{equation}
 F = \frac{(1 -  |\Gamma_l|^2)^{1/2}}{(1 - \Gamma_a\Gamma_lL)}
\end{equation}
$L$ is the loss factor of the cable or attenuator and

\begin{equation}
 \Gamma_a = \Gamma_{aa}exp(i\omega\tau)
\end{equation}
where $\tau$ is the 2-way cable delay and $\Gamma_{aa}$ is the antenna reflection coefficient without the cable or attenuator.

\section{Model for choke balun loss}
 
    The "choke" balun shown in Figure 1 is modelled by the circuit shown in Figure 2. In this model twice the ferrite impedance appears
in parallel across the balanced antenna terminals so the reflection coefficient measured at the unbalanced port, $\Gamma$,
is

\begin{equation}
 \Gamma = Lexp(i\omega\tau)\Gamma_b
\end{equation}
where $\Gamma_b$ is the reflection coefficient of the antenna impedance $Z_a$, in parallel with twice the choke impedance, $Z_f/2$,

\begin{equation}
 Z_b = (1/Z_f + 1/Za)^{-1}
\end{equation}

\begin{equation}
 \Gamma_b = \frac{(Z_b-50)}{(Z_b+50)}
\end{equation}
Since the noise power from impedance $Z$ is proportional to its temperature times the real part of the impedance
the loss factor is

\begin{equation}
 L  = \frac{ Re(Z_a)|Z_f|^2}{(Re(Z_a)|Zf|^2+Re(Zf)|Z_a|^2)}
\end{equation}
 
\section{Example of calibrated sky spectrum measurement from 55 to 110 MHz}

    A short observation of the sky spectrum was made with a broadband dipole and a 3-position switched spectrometer
at a remote site away from strong TV and RF radio signals
to test the calibration method. The switched spectrometer made the measurements of the following:

 1]  50 ohm resistor heated to 100 K above ambient

 2] An open cable about 3m in length 

 3] Sky noise at a remote site

VNA measurements were made of the following:

 1] Antenna through the balun at the remote site

 2] Input of the LNA through the 3-position switch

 3] Ferrite balun impedance and back-to-back loss

 4] Open cable used to derive the LNA noise parameters

    Measurements of the heated resistor were used to calibrate the internal noise diode in flatness and absolute
scale. To remove any uncertainty in the temperature of this "hot" the measurement was checked using a HP346C 
noise source with calibration traceable to NIST.

    The wideband dipole used for the test
is based on the "Fourpoint" design of  {\textit  {Suh et al.}} [2004]. This antenna which was made of aluminum rod 
elements instead of panels is shown in Figure 3. Figure 4 shows a block diagram of the system used to measure the sky noise
spectrum which includes the antenna, balun, 3-position switch, LNA and spectrometer. It also shows a noise source used to inject
added noise below 55 MHz which conditions the analog to digital converter (ADC). This "out of band" conditioning improves the linearity
and dynamic range of the ADC.
 
    Figure 5 shows the measured reflection magnitude and phase
along with a Fourier series least square fit to the data to reduce the noise in the VNA data and provide accurate 
interpolation between the measurement.
 
    Figure 6  shows the measured reflection coefficient of the LNA. 
In this case the value of fitting a Fourier series is clearly evident as the VNA measurements of the LNA were make with a spacing 
of 10 MHz.
 
    Figure 7 shows the spectrum of the calibrated 3-position switched spectrometer connected to the open cable
with  2-way delay of 29 ns and 0.7 dB loss. The solid curve is the spectrum derived from the best fit LNA noise parameters
shown in Figure 8. The frequency dependence of the noise parameters were constrained to a constant plus a
slope. At 80 MHz the values of $T_u$,$T_c$,$T_s$ were 35, 9, and 10 respectively.

    Figures 9 and 10 show the calibrated sky spectrum for 2 sequential observations taken at the remote site 
in West Forks, Maine on 8 July 2011 at 17 UT. 
In the first the antenna was connected directly to the 3-position switch and in the second the antenna
was connected through a 6 dB attenuator.
These calibrated spectra were derived using equation (18) or equation (21), in the case of the use of the added attenuation,
followed by equation (19) for the correction of the antenna loss factor. The balun loss factor was calculated using equation
(25) assuming a ferrite choke impedance of 400 ohms for which the circuit model 
gives a back-to-back loss of 0.52 dB for back-to-back baluns consistent with 
the measured value. A loss factor of 0.93 was assumed for the ground loss of 7\% based on the results of an electromagnetic
simulation of the antenna and ground plane. The overall loss factor is the product of the balun and ground loss factors. The antenna
was assumed to have no resistive loss. 

\section{Determination of spectral index}

An estimate of the spectral index is obtained by finding the the function,

\begin{equation}
 T_{sky}(f) = T_{sky}(f0)(f/f0)^{-sp}
\end{equation}
where $T_{sky}(f)$ is the sky temperature as a function of frequency $f$, $sp$ is the spectral index and $f0$ is the "reference" frequency,
that fits the calibrated sky noise spectrum using weighted least squares for which portions of
the spectrum with added interference are given zero weight. The best fit spectral indices using the calibrated 
sky spectra shown in Figures 9 and 10 were 2.50 +/- 0.05  and 2.53 +/- 0.05 respectively. These values are in good agreement with
measurements made from 100 to 200 MHz by {\textit  {Rogers and Bowman}} [2008] in which the antenna was connected to the antenna
via a long cable so that correlated noise terms in equation (19) are fitted and removed. The disadvantage
of this method is that a long cable has to be used to provide the rapid change of phase of $\Gamma_a$ which results in added
loss. In addition the "ripples" in the measured spectrum are assumed to arise only the mismatch and cannot be separated from 
fine scale structure which might be present in the sky spectrum. In this method the antenna reflection coefficient  
was measured by injecting added noise which was then reflected by the antenna and measured by the spectrometer. But 
the antenna reflection coefficient could have been measured with a VNA.

\section{Sensitivity of calibrated sky temperature to VNA measurement errors}

The sensitivity to measurement errors in the ancillary measurement of antenna and LNA reflection 
coefficient, balun ferrite impedance and ambient temperature was determined by
using equation (19) to generate noiseless "reference" data assuming a sky temperature of 1500 K at 80 MHz
and a spectral index of -2.5. New data was then 
generated by perturbing the ancillary VNA data, first by 1\% in amplitude, and then by 0.01 radians in phase. A test of
the sensitivity to the ambient temperature was also made by
changing the assumed ambient temperature by 1 K. These two data sets were then processed to 
obtain the calibrated sky temperature and after removing the best fit spectral index their rms differences computed. 
The results with an attenuator of 10 dB placed between the antenna and the 3-position switch are given in Table 1.

\begin{table}
\caption{Sensitivity to changes in ancillary measurements\tablenotemark{a}}
\centering
\begin{tabular}{l l l}
\hline
Parameter changed  & rms (K) & rms2 (K)  \\
\hline
  $\Gamma_a$ amplitude & 1.8 & 0.09   \\
  $\Gamma_a$ phase     & 0.25 & 0.06  \\
  $\Gamma_l$ amplitude & 0.13 & 0.06 \\
  $\Gamma_l$ phase     & 0.19 & 0.04  \\
  $Z_b$     amplitude & 0.38 & 0.09 \\
  $Z_b$     phase     & 0.004 & 0.001  \\
  $T_{amb}$ temperature & 1.2 & 0.8  \\
\hline
\end{tabular}
\tablenotetext{a}{rms2 is for antenna with -20 dB reflection coefficient}
\end{table}

For long integrations for which the systematics still dominate an attenuation of about 10 dB
is optimum for the high sky noise temperature at 80 MHz.
Using more attenuation makes little difference to the sensitivity to errors in the
ancillary measurements and using no attenuator increases the sensitivity to error in
the reflection coefficients by about a factor of 2. It is noted that the most critical measurement
is that of the antenna impedance. However the sensitivity to antenna impedance can be reduced
if an antenna with a better match over the wide frequency range can obtained. For example a large reduction
in sensitivity could be obtained with an antenna whose the reflection coefficient is below -20 dB from 55 to 110 MHz
as shown by the numbers in the last column of Table 1. A reduction of the sensitivity to antenna reflection
coefficient by a factor of about 3 can be obtained with the "Fourpoint" used in the test if the frequency range is
limited to 60 to 90 MHz. We note that precision impedance analyzers are available which may achieve higher accuracy
than a VNA in the frequency range below 110 MHz.

\section{Conclusion}

The accurate calibration method described has the potential to make wideband sky noise spectral measurements
with greatly reduced level of systematics. Simulations of performance look encouraging but emphasize the very high
accuracy needed in the measurements of reflection coefficient. 

\begin{acknowledgments}
   We thank Delani Cele of Ithaca College, New York who made the antenna and balun VNA measurements and 
helped with the set-up of the dipole at West Forks, Maine 
while he was spent the summer of 2011 at Haystack Observatory under the NSF supported  Research Experiences for Undergaduates (REU) program. 
\end{acknowledgments}

\end{article}

\pagebreak

\begin{figure}[b]
\noindent\includegraphics[width=39pc]{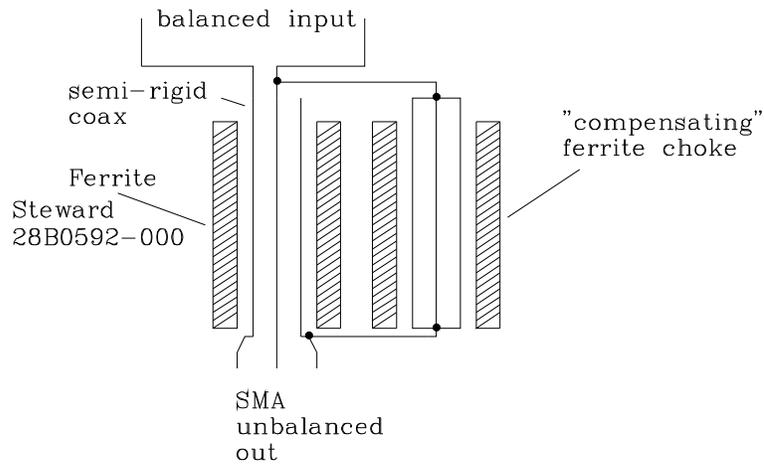}
\caption{Ferrite choke balun}
\end{figure}
%\end{document}

\begin{figure}
\noindent\includegraphics[width=39pc]{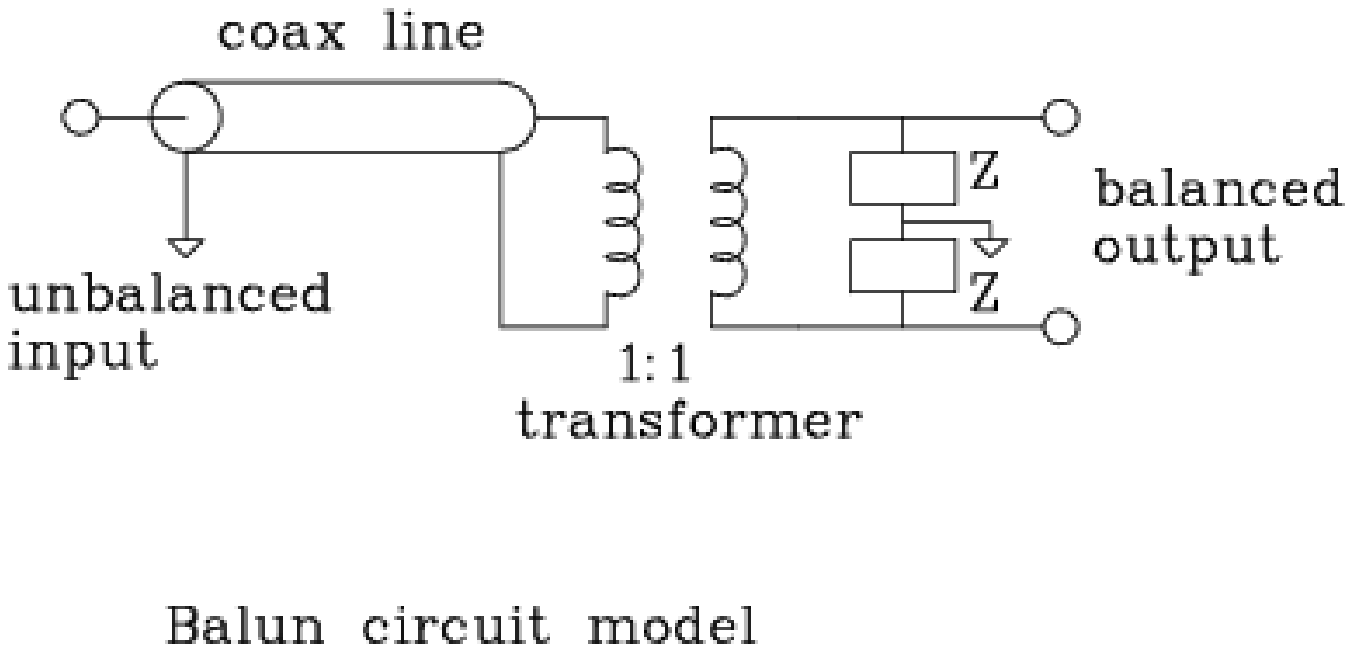}
\caption{Circuit model of ferrite choke balun}
\end{figure}
%\end{document}

\begin{figure}
\noindent\includegraphics[width=39pc]{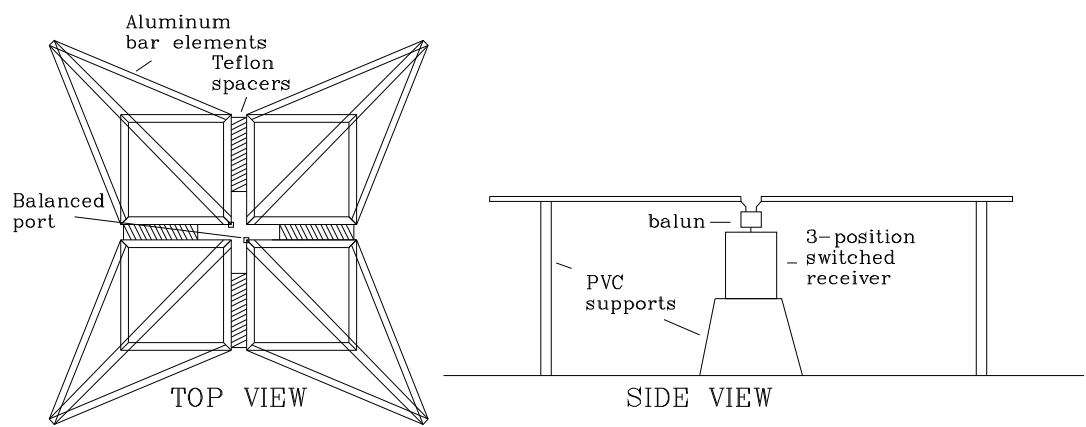}
\caption{Fourpoint antenna $0.43\lambda$ tip-to-tip dipole $0.2\lambda$ above ground at 55 MHz}
\end{figure}
%\end{document}

\begin{figure}
\noindent\includegraphics[width=39pc]{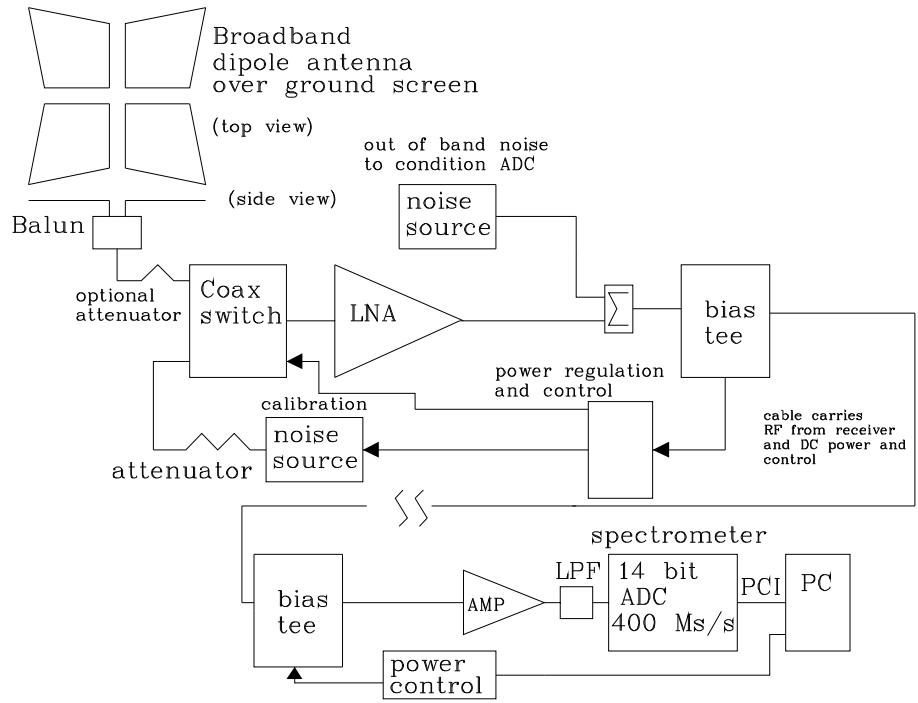}
\caption{Block diagram of complete system for sky noise measurements}
\end{figure}

\begin{figure}
\noindent\includegraphics[width=39pc]{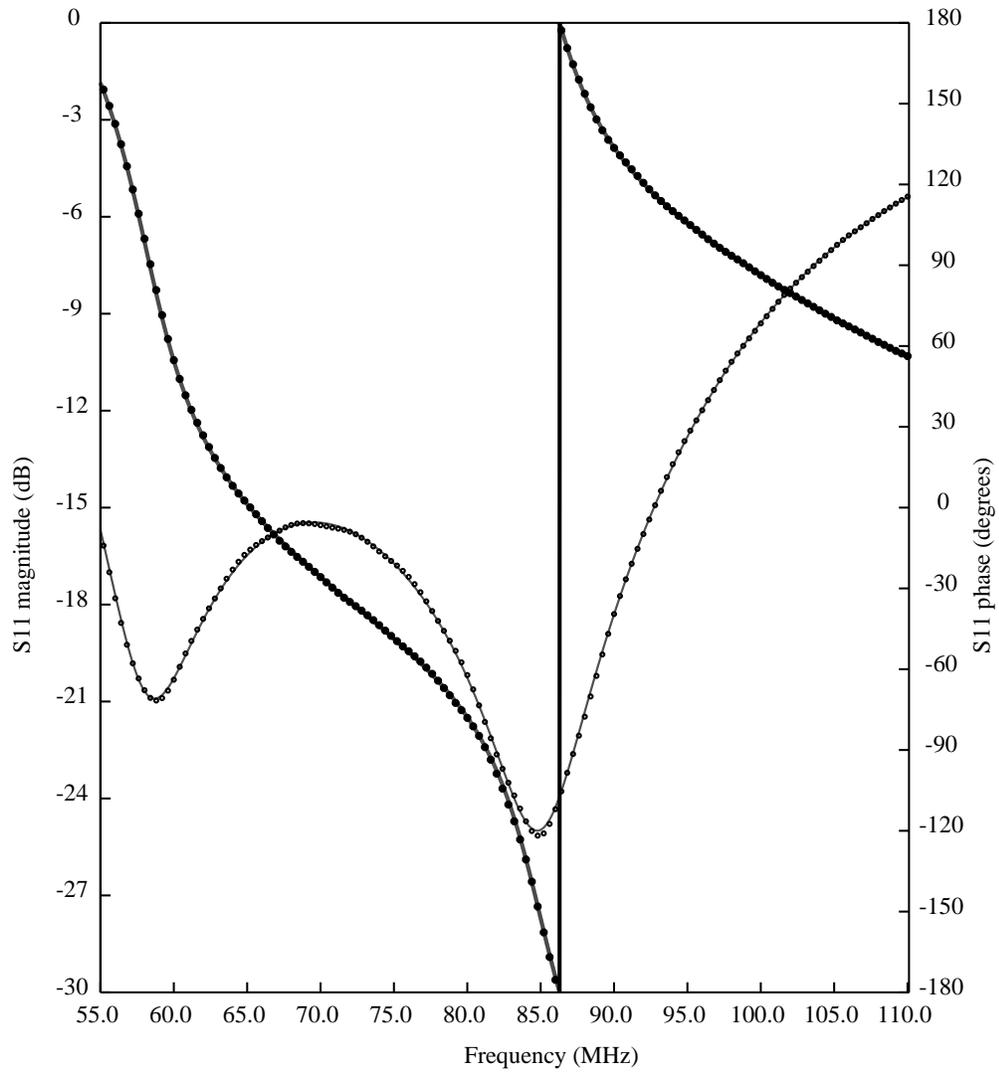}
\caption{Antenna reflection coefficient}
\end{figure}
%\end{document}

\begin{figure}
\noindent\includegraphics[width=39pc]{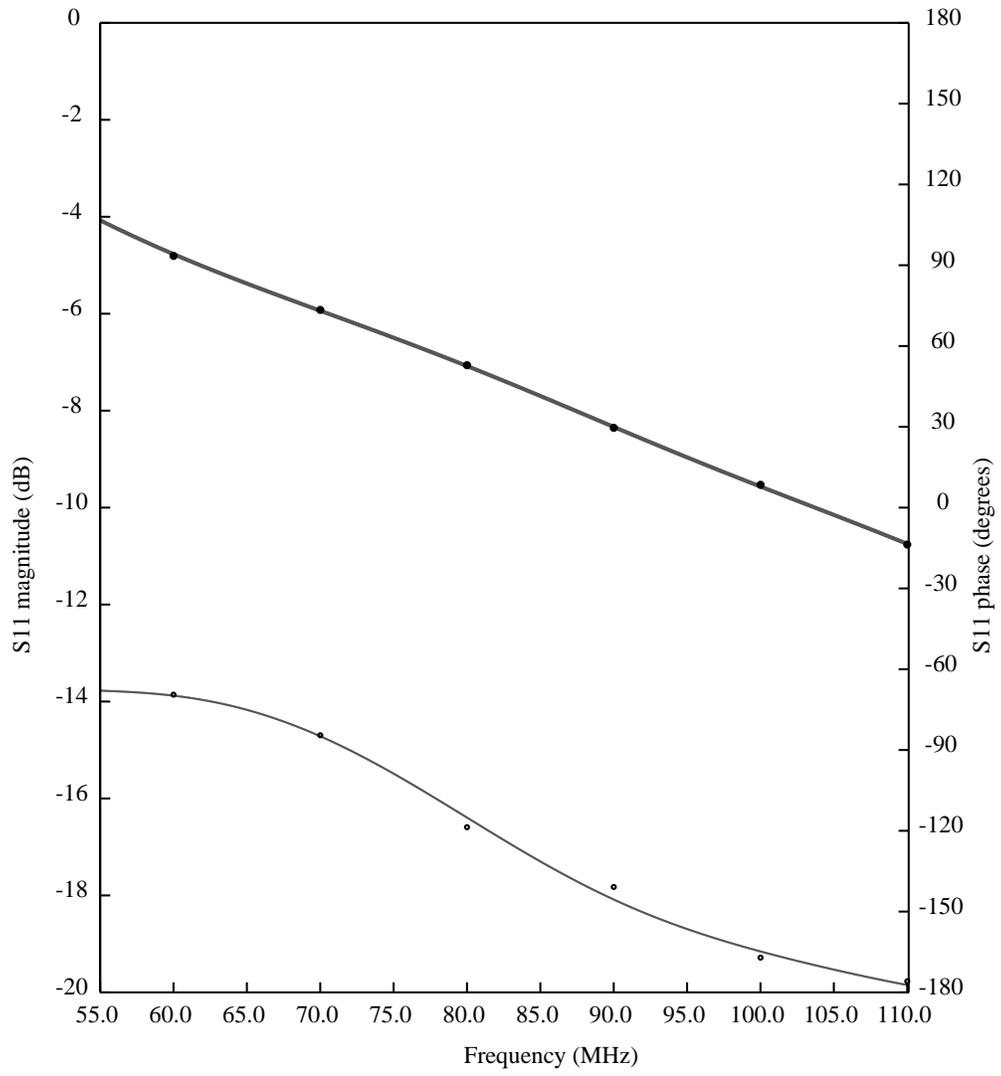}
\caption{LNA reflection coefficient}
\end{figure}
%\end{document}

\begin{figure}
\noindent\includegraphics[width=39pc]{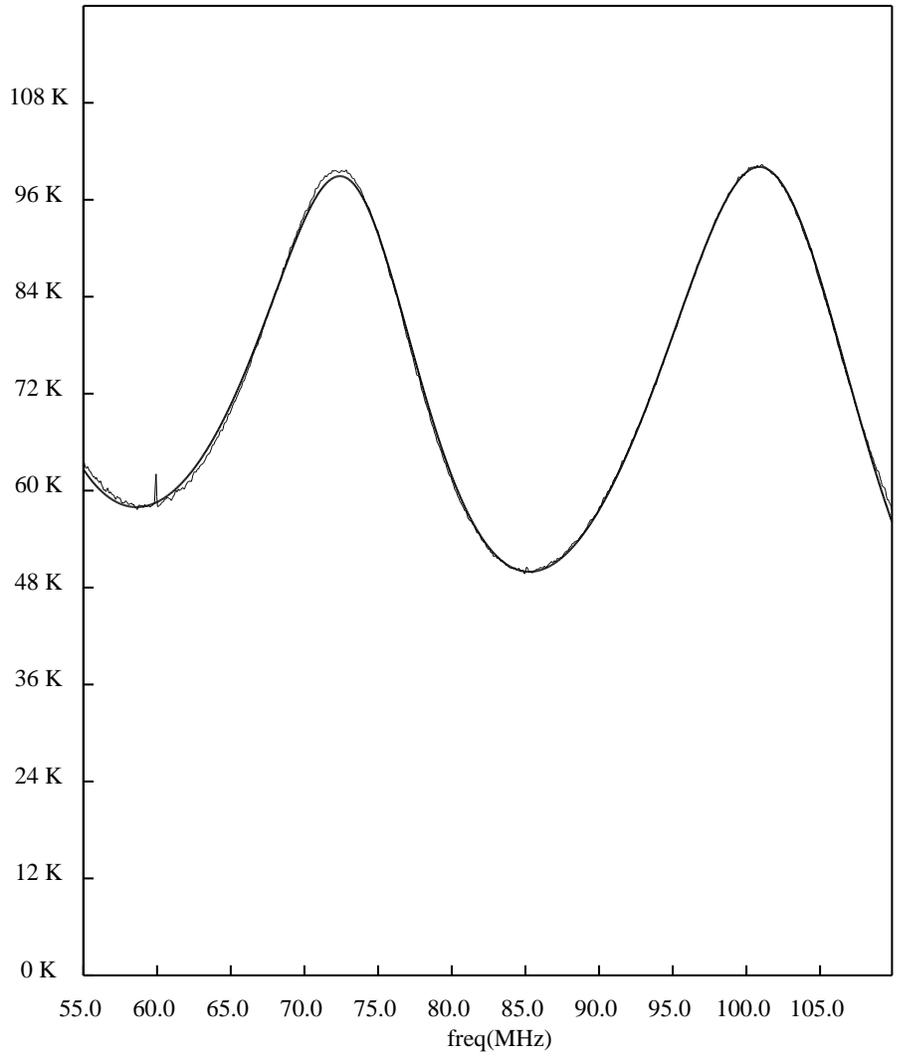}
\caption{Spectrum of open cable}
\end{figure}
%\end{document}

\begin{figure}
\noindent\includegraphics[width=39pc]{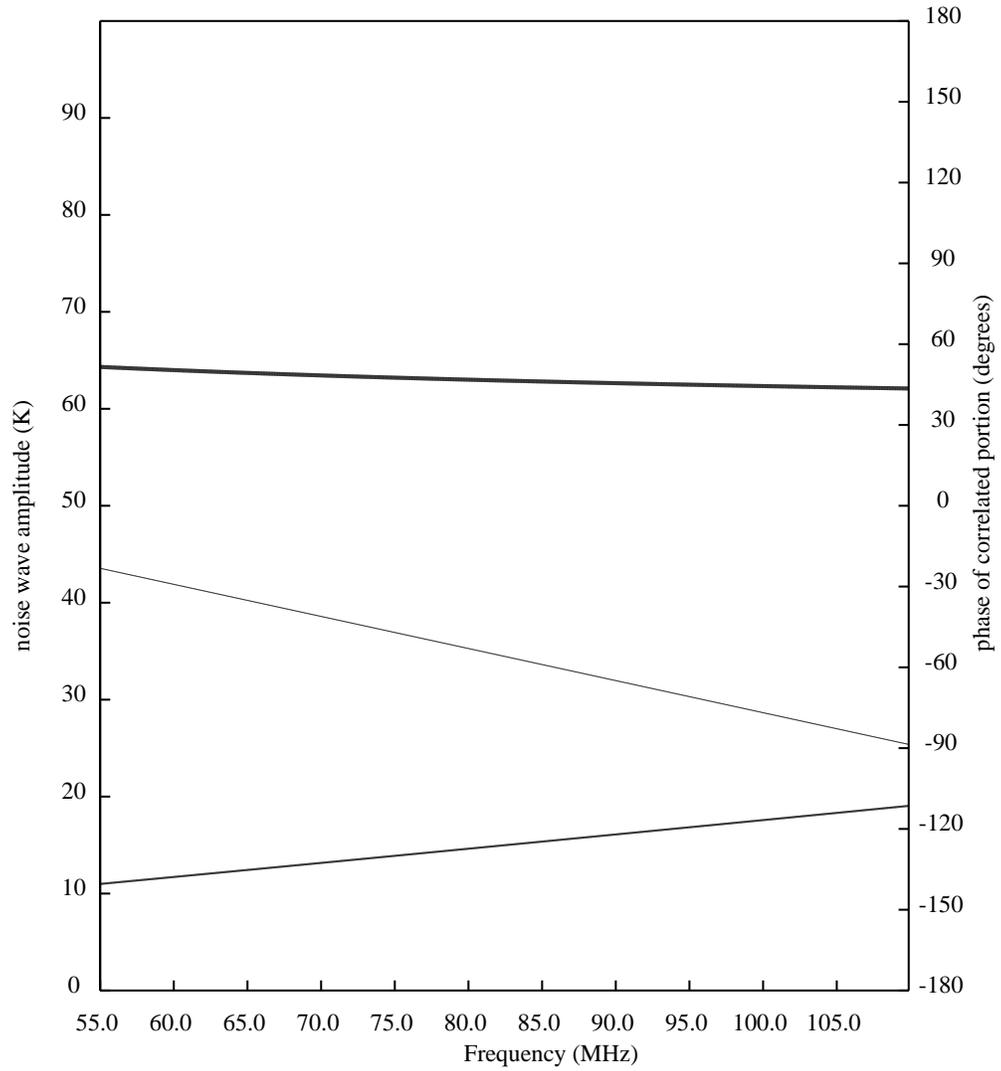}
\caption{LNA noise parameters. Top curve is the phase of the correlated noise. The middle curve is uncorrelated noise and the bottom curve
is the magnitude of the correlated noise}
\end{figure}
%\end{document}

\begin{figure}
\noindent\includegraphics[width=39pc]{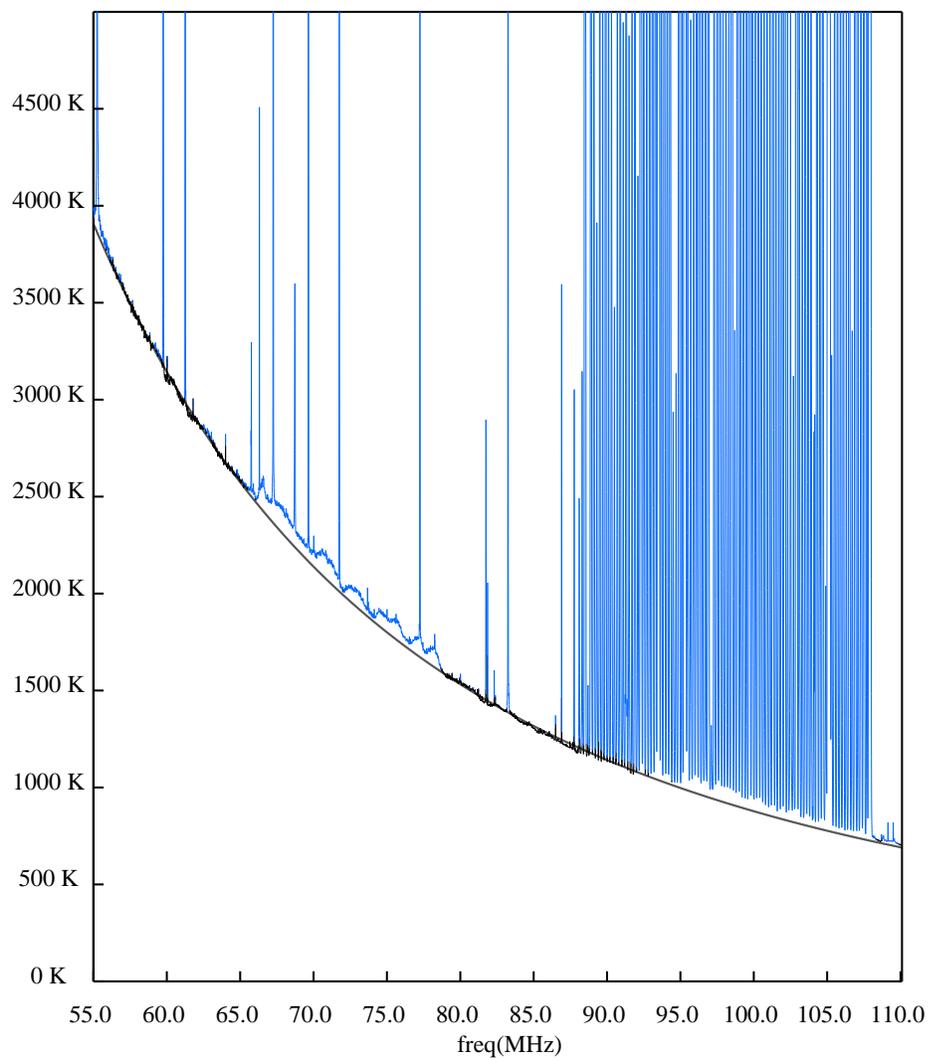}
\caption{Calibrated sky noise spectrum without attenuation}
\end{figure}
%\end{document}

\begin{figure}
\noindent\includegraphics[width=39pc]{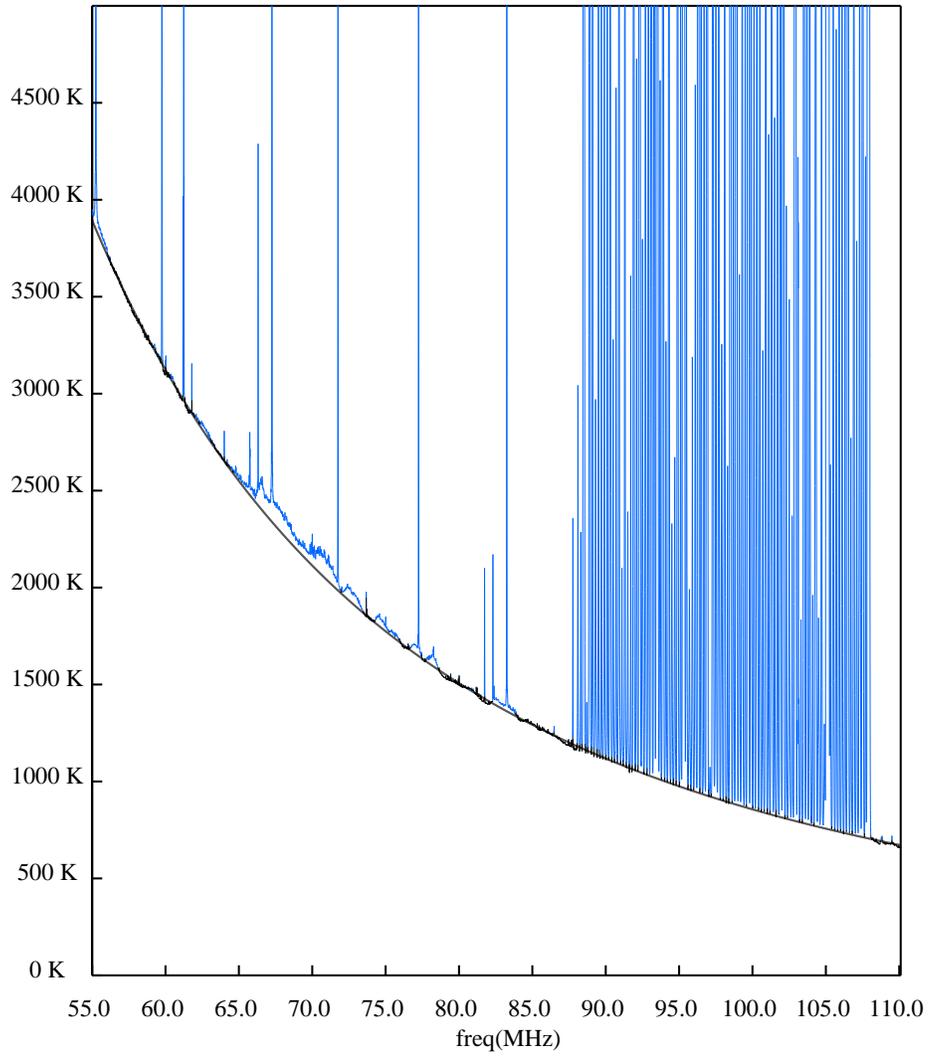}
\caption{Calibrated sky noise spectrum with 6 dB attenuation}
\end{figure}
\end{document}